\documentclass[prl,twocolumn,showpacs,preprintnumbers,amsmath,amssymb,tightenlines,epsfig,superscriptaddress]{revtex4}
\usepackage{graphicx}
\usepackage{color}
\usepackage[normalem]{ulem}
\usepackage[percent]{overpic}
\newcommand{\bra}[1]{\langle\,{#1}\, |}
\newcommand{\ket}[1]{|\,{#1}\,\rangle}

\newcommand{\rvec}[1]{{\mathbf{r}}}

\newcommand{\ssection}[1]{{\noi  \it #1:}}

\newcommand{\sub}[2]{{#1}_{\mbox{\!\! \scriptsize #2}}}

\def\noi{\noindent}
\def\beq{\begin{equation}}
\def\eeq{\end{equation}}

\newcommand{\fref}[1]{Fig.~\ref{#1}}
\newcommand{\frefp}[2]{Fig.~\ref{#1}~(#2)}

\newcommand{\eref}[1]{Eq.~(\ref{#1})}

\newcommand{\cref}[1]{chapter~\ref{#1}}

\newcommand{\Cref}[1]{Chapter~\ref{#1}}

\newcommand{\bref}[1]{(\ref{#1})}

\usepackage{ulem}  
\usepackage{bbold}
\usepackage{braket}
\usepackage{float}

\begin{document}

\title{Quantum transport enabled by non-adiabatic transitions}
\author{Ajith Ramachandran}
\affiliation{Department of Physics, Indian Institute of Science Education and Research, Bhopal, Madhya Pradesh 462 066, India}
\affiliation{Department of Physics, Christ College, Irinjalakuda, Kerala 680125, India}
\author{Alexander Eisfeld}
\affiliation{Max Planck Institute for the Physics of Complex Systems, N\"othnitzer Stra{\ss}e 38, D-01187 Dresden, Germany}
\author{Sebastian W\"uster}
\email{sebastian@iiserb.ac.in}
\affiliation{Department of Physics, Indian Institute of Science Education and Research, Bhopal, Madhya Pradesh 462 066, India}
\author{Jan-Michael Rost}
\affiliation{Max Planck Institute for the Physics of Complex Systems, N\"othnitzer Stra{\ss}e 38, D-01187 Dresden, Germany}
\begin{abstract}
Quantum transport of charge or energy in networks with discrete sites is central to diverse quantum technologies, from molecular electronics to light harvesting and {quantum opto-mechanical metamaterials}. A one dimensional network can be viewed as waveguide. We show that if such waveguide is hybridised with a control unit that contains a few sites, then transmission through the waveguide depends sensitively on the motion of the sites in the control unit.
Together, the hybrid waveguide and its control-unit form a Fano-Anderson chain whose Born-Oppenheimer surfaces inherit characteristics from both components: A bandstructure from the waveguide and potential energy steps as a function of site coordinates from the control-unit. Using time-dependent quantum wave packets, we reveal conditions under which the hybrid structure becomes transmissive only if the control unit contains mobile sites that induce non-adiabatic transitions between the surfaces. Hence, our approach provides functional synthetic Born-Oppenheimer surfaces for hybrid quantum technologies combining mechanic and excitonic elements, and has possible applications such as {switching and} temperature sensing.
\end{abstract} 
\maketitle

\ssection{Introduction} Excitation transport on networks of monomeric quantum units is central
 in diverse physical settings ranging from photosynthesis \cite{van2000photosynthetic} over quantum dot arrays \cite{kagan1996long}, molecular aggregates \cite{saikin:excitonreview}, conjugated polymers \cite{fratini2020charge,pace2019intrinsically} and {quantum opto-mechanics} \cite{lu2014topological,galisteo2011self,douglas2015quantum,safavi:optomecharray,Ludwig:QMBoptomecharray,Schmidt:Diracoptomecharray} to single-molecule devices \cite{qiu2004vibronic,mitra2004phonon,sowa2017vibrational,xin2019concepts}. Motion of constituting elements, referred to as sites in the following, is inevitable in most of these systems, but often regarded as obstructing transport \cite{mannouch2018ultra,schneider2013dissipative}. Usually this motion is considered in the adiabatic Born-Oppenheimer (BO) approximation for the electronic ground state. 

Going beyond this approximation, we show that non-adiabaticity can be turned into a feature to control transport along a regular chain that acts as waveguide for excitation,
if the chain is coupled to a control unit (CU) with a few sites which can indeed move. This combines a Fano-Anderson chain  \cite{miroshnichenko2010fano}, with a vibrational mode.
The resulting adiabatic (Born-Oppenheimer) surfaces show hybrid characteristics of both components: regular, dense bands as expected for the long, ordered chain are modified by a series of potential energy steps induced by the slanting BO surface of the CU.  
When immobile, the system exhibits Fano resonances \cite{fano1935sullo,fano1961effects,miroshnichenko2010fano} that cause total reflection of guided wavepackets at certain energies. At these energies, the CU can fully prohibit transmission in the chain \cite{miroshnichenko2005engineering} if motion along the BO surfaces is adiabatic.
 However, the potential steps introduce the possibility to traverse them diabatically for suitable motion of the CU sites. Hence, motion can induce
 transport rather than blocking it. 

\begin{figure}[htb]
\includegraphics[width=0.99\columnwidth]{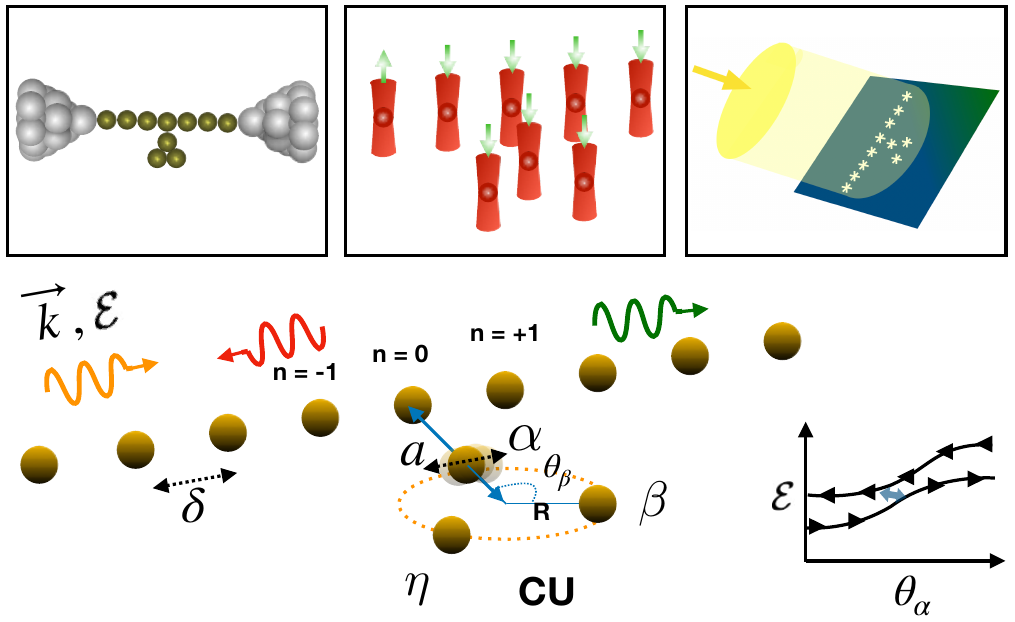}
\caption{\label{sketch} Sketch of the Fano-Anderson chain with vibrating elements. $N$ sites are arranged in a linear chain while three sites $\alpha,~\beta,~\eta$ form a control-unit CU with geometry as discussed in the text. Site $\alpha$ may vibrate along its angular coordinate $\theta_\alpha$ on the ring as shown. A single excitation wavepacket near energy ${\cal E}$ and central wavenumber $\mathbf{k}$ approaching the control-unit from the far left (yellow) is usually reflected (red) but may be transmitted (green) due to non-adiabatic transitions only.
The latter are sketched on the right, as transitions between eigenstates localised on the left (right) side of the CU, as indicated by $\blacktriangleleft$ ($\blacktriangleright$).
The panels on the top visualize key platforms described by our abstract model, (left) molecular wires, (middle) cold atom lattices, (right) quantum opto-mechanical arrays. }
\label{geometry}
\end{figure}
We will illustrate the general approach with a trimer as CU. The ensuing model could be realized with molecular wires of conjugated polymers \cite{schwartz2003conjugated}, Rydberg aggregates \cite{wuester:review} or {opto-mechanical elements} \cite{miroshnichenko2010fano}. Cold atoms would provide a controllable platform to confirm our predictions, while the link between motion and transport could enable temperature sensors in molecular electronics \cite{schwartz2003conjugated}. 

\ssection{Transport model} We consider a linear chain of $N$ discrete sites (monomers) with an attached CU consisting of three sites confined to a circle of radius $R$, as shown in \fref{geometry}. This setup constitutes a simple geometry to introduce controlled non-adiabatic dynamics. We label the chain sites by integers and those of the CU by $\alpha$, $\beta$ and $\eta$. Each site forms a two-level system with states $\ket{g}$, $\ket{e}$, the ground and excited state, or empty and occupied state. The number of excitations is conserved and we focus on the single excitation Hilbert space, spanned by $\ket{\pi_n} = \ket{ggg\dots e \dots ggg}$ in which only the $n^{\text{th}}$ site is excited. For simplicity, {\it all} sites are assumed to be immobile, except the site $\alpha$ of the control-unit which vibrates in a harmonic potential with frequency $\omega$ and effective mass $M$. The positions $\boldsymbol{r}_n$ of the sites can be grouped into a $3(N+3)$ component vector $\boldsymbol{R}=\{\boldsymbol{r}_n\}$. Then, the Hamiltonian ($\hbar=1$) describing excitation, long range interactions and motion reads
\begin{align}
\hat{H} &= - \frac{\boldsymbol{\nabla}_{\boldsymbol{r}_\alpha}^2}{2M} + V(\boldsymbol{r}_\alpha)  +\sub{\hat{H}}{el}(\boldsymbol{R}),
\label{ham}\\
\sub{\hat{H}}{el}(\boldsymbol{R}) &= \sum_{n,m}^{N+3}W_{nm}({\boldsymbol{r}_n} , {\boldsymbol{r}_m})\ket{\pi_n}\bra{\pi_m}, 
\label{ham_e}
\end{align}
where $V(\boldsymbol{r}_\alpha)$ is a harmonic potential. For $n=m$, $W_{nn}=E_n$ is the onsite energy of monomer $n$.
For $n\ne m$, the interaction $W_{nm}$ allows the excitation to migrate. To be specific, we take 
$W_{nm}=-C_3/|\boldsymbol{r}_n-\boldsymbol{r}_m|^3$, which could be realized by dipole-dipole interactions \cite{robicheaux2004simulation,wuester:review}. Interactions on the control-unit depend on the relative angles $\theta_{\gamma\delta}=\theta_\delta-\theta_\gamma$, where $\theta_\gamma$ ($ \gamma \in \{\alpha,\beta,\eta\}$) is the 2D polar angle of monomer $\gamma$ on the ring, as shown in \fref{geometry}. 
To reduce complexity, we only consider nearest neighbour interactions, such that $W_{nm}=\delta_{nm} J$ on the chain, which is coupled to the CU via interactions $W_{0\alpha}=W_{0\alpha}(\theta_\alpha)$ between sites $0$ and $\alpha$.
Now, $\sub{\hat{H}}{el}(\boldsymbol{R})$ non-trivially depends on all positions, such that the electronic and vibrational degrees of freedom are coupled and the vibration of  site $\alpha$ can significantly affect excitation transport. 
While we keep the model simple to focus on the essential physics, the effects reported persist also for long-range interactions between sites. Realizing a site $\alpha$ which is much more mobile than the main chain is feasible with many of the platforms listed in the introduction.

\begin{figure}[htb]
\includegraphics[width=\columnwidth]{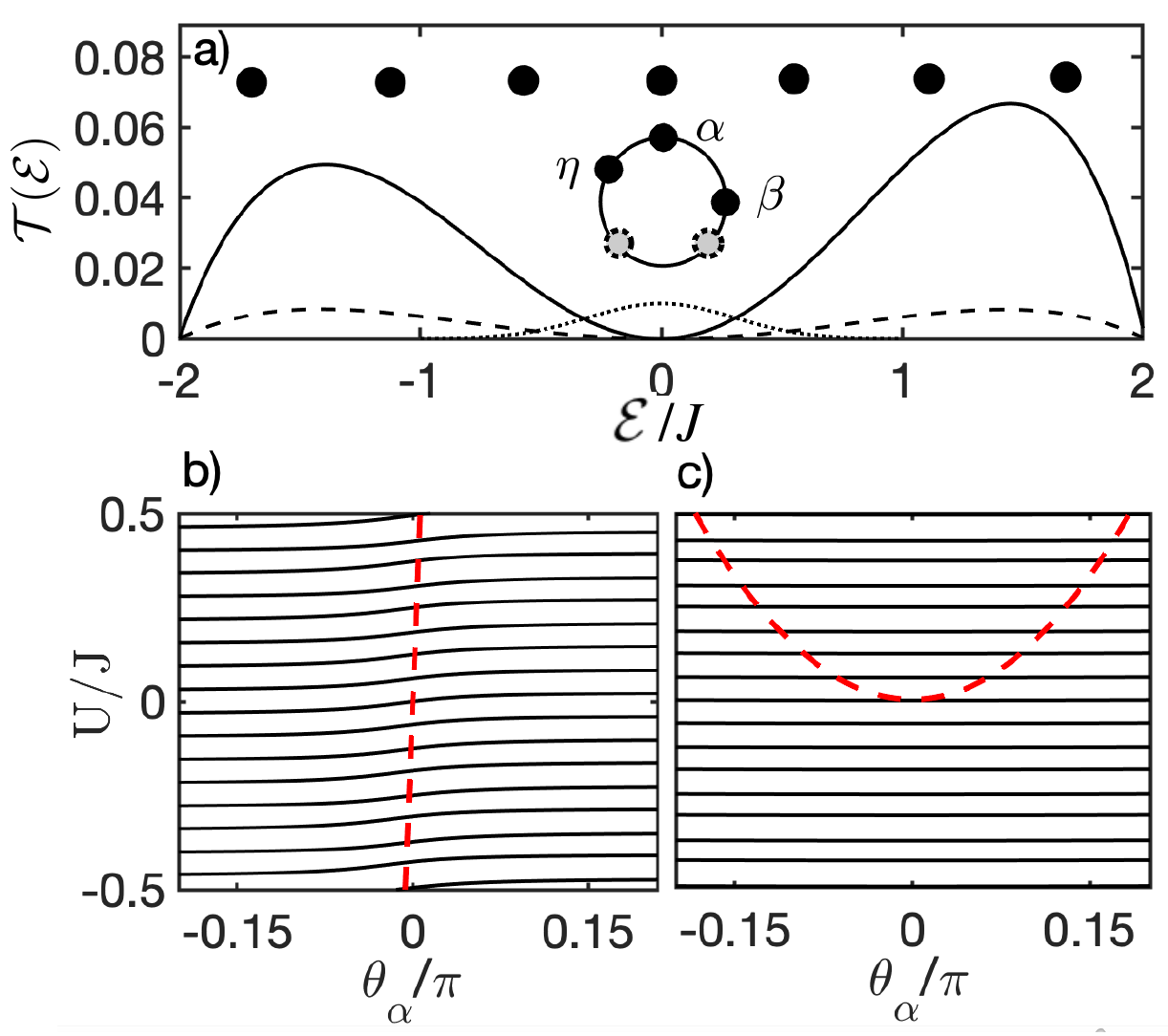}
\caption{Hybrid system, transmission and energies. (a) Very small transmission ${\cal T}$ past the control-unit if sites are immobile, as function of incoming energy $\cal E$ of the excitation, for the two configurations shown. The solid (dashed) curve corresponds to the configuration with (without) a slope on the BO surfaces. Solid (dotted) circles indicate monomer positions. The dotted Gaussian line is the scaled energy distribution of width $\sigma_{\cal E}$ for the initial electronic wavepacket arriving from the left of the chain.
(b) Spectrum (black lines) of the electronic Hamiltonian $\sub{\hat{H}}{el}$ in \eref{ham_e} for a chain of $N=20$ sites that is hybridized with a trimer control-unit, at solid circles in (a). We consider monomer $\alpha$ mobile on a ring, with polar angle $\theta_\alpha$. The red dashed line is one energy $U_j^{(C)}$ of the trimer control-unit without chain. For other parameters and geometry, see {footnote} \cite{par_main}. (c) The same for the dotted {monomer} positions in (a).
}
\label{spectrum}
\end{figure}

\ssection{Hybridising waveguide and control unit} We now show that the chosen geometry yields an amalgamation of
the spectra characteristic of a regular chain (band structure of flat BO surfaces) with those of the control unit (slanting BO surfaces of discrete eigenvalues). Solving the electronic eigenvalue problem for the control-unit only, $\sub{\hat{H}}{el}^C(\theta_\alpha) \ket{\psi^C_j(\theta_\alpha)} = U_j^C(\theta_\alpha)\ket{\psi^C_j(\theta_\alpha)}$ provides three adiabatic BO potential energy surfaces $U_j^C(\theta_\alpha)$ for the motion of the sites on the circle, with
 $\sub{\hat{H}}{el}^C(\theta_\alpha)$ from \bref{ham_e} with $n,m$ restricted to $\{\alpha,\beta,\eta\}$. 
 In accordance with our minimal model introduced above, we restrict motion of the sites further to site $\alpha$ with angle $\theta_\alpha$ only. The fixed locations of the other sites are shown as ($\bullet$) in \frefp{spectrum}{a}, for two geometries. A chain without control unit has a dense band of flat states covering the energy range from $-2J$ to $2J$. The colored dashed line in \frefp{spectrum}{b} shows one energy of the isolated control unit $U_j^C(\theta_\alpha)$ for $\theta_{\beta}=\pi/2$ and $\theta_{\eta}=5\pi/3$. It passes through the energy band of the main chain.

Now, when coupling the CU with $W_{0\alpha}(\theta_\alpha)\neq 0$ to the linear chain, its band structure inherits CU energies as a series of slanting steps, as shown in \frefp{spectrum}{b}. 
These can be regarded as relatively wide avoided crossings, that would become narrower with steeper steps for smaller $W_{0\alpha}$. 

In contrast, for an alternative CU configuration, in which the sites $\beta$ and $\eta$ are further away from site $\alpha$, as shown by gray dashed circles in \frefp{spectrum}{a}, interactions between site $\alpha$ and sites $\beta$, $\eta$ become less relevant.  As a result, even the bandstructure of the hybrid system contains almost flat energy surfaces without steps over the relevant range of motion for $\theta_\alpha$, see \frefp{spectrum}{c}.  We will now show how the potential steps induced by the CU can be exploited to control quantum transport across the main chain. We thus describe a scheme to taylor functional Born-Oppenheimer surfaces, by merging desirable properties from two different structures.

\ssection{Quantum dynamics} 
Within the total quantum state $\ket{\Psi(t)} = \sum_{n=1}^{N+3} \phi_n(\theta_\alpha,t)\ket{\pi_n}$, the wavefunction corresponding to the motion of site $\alpha$ (with angle $\theta_\alpha$) in the electronic state $\ket{\pi_n}$ is $\phi_n(\theta_\alpha,t)$. Its dynamics follows from \bref{ham_e} as:
\begin{align}
\label{tdse}
i\frac{\partial}{\partial t}\phi_n(\theta_\alpha) =& -  \frac{1}{2M R^2}\frac{\partial^2 }{\partial \theta_\alpha^2} \phi_n(\theta_\alpha) + \sum_{m}W_{nm}\phi_m(\theta_\alpha) \\&+ \frac{1}{2}M\omega^2 R^2 \theta_\alpha^2 \phi_n(\theta_\alpha). \nonumber
\end{align}

The excitation coming in from the far left of the linear chain forms a Gaussian wave packet in site index $n$, while the initial vibrational quantum number is $\nu$. Hence,
 $\phi_n{(\theta_\alpha,t=0)} = {\cal N} \exp{[ikn-(n-n_0)^2/(2\sigma^2)]}\varphi_\nu{(\theta_{\alpha}-\theta_{\alpha_0})}$, where ${\cal N} $ is a normalization constant, $n_0$ is the initial mean site of the wavepacket and $\sigma$ the corresponding width, $\theta_{\alpha_0}{=0}$ is the mean angular position of monomer $\alpha$ and $\varphi_\nu$ the $\nu$'th harmonic oscillator state. The central wavenumber $k$ is linked to the incoming energy ${\cal E}$ via $k=\cos^{-1}({\cal E}/2J)$. 

We use \eref{tdse} for propagation. To analyze non-adiabatic transitions, in addition we express the total state in the adiabatic basis $\ket{\Psi(t)} = \sum_k \tilde{\phi}_k(\theta_\alpha,t)\ket{\psi_k(\theta_\alpha)}$, with $\sub{\hat{H}}{el}\ket{\psi_k(\theta_\alpha)}=U_k(\theta_\alpha)\ket{\psi_k(\theta_\alpha)}$. The vibrational wavefunction on a specific BO surface, $\tilde{\phi}_k(\theta_\alpha,t)$, obeys 
\begin{align} \label{tdse_ad} 
i\frac{\partial}{\partial t}\tilde{\phi}_k(\theta_\alpha)= &\left[ -  \frac{1}{2MR^2}\frac{\partial^2 }{\partial \theta_\alpha^2}   +U_k(\theta_\alpha)\right]\tilde{\phi}_k(\theta_\alpha)  \\&+ \sum_l D_{kl}(\theta_\alpha)\tilde{\phi}_l(\theta_\alpha), \nonumber
\end{align}
where the non-adiabatic coupling terms (NAC) $D_{kl}(\theta_\alpha)$ are discussed in detail in the the Supplemental Material (section I). While transmission past a CU with internal vibrations can also be treated as a multi-channel scattering problem \cite{ramachandran:multichannel}, our time-dependent approach enables us to directly connect transmission to non-adiabatic transitions.

\ssection{Non-adiabatic transmission} 
The waveguide with CU has been designed such that transmission past the CU is nearly completely suppressed in the immobile case, but can be made likely through non-adiabatic transitions induced by motion. 
Without motion, the transmission coefficient ${\cal T}({\cal E})$ for passing the CU at energy ${\cal E}$ can be calculated with a transfer matrix method \cite{miroshnichenko2005engineering,miroshnichenko2010fano}.
In \frefp{spectrum}{a} ${\cal T}({\cal E})$ is shown for a specific set \cite{par_main} of parameters.
Transmission vanishes for energies ${\cal E}$ resonant with an eigenenergy $U_j^{(C)}$ of the isolated control-unit, i.e.~${\cal T}=0$ for ${\cal E}\approx U_j^{(C)}$.
We have chosen the on-site energies $E_n$ of the CU such that resonance occurs at ${\cal E}=0$.
The coupling $W_{0\alpha}$ between chain and CU has been adjusted such that the width of the resonance would be much wider than the exciton bandwidth $-2J<{\cal E}<2J$.  As a result, transmission is strongly suppressed $({\cal T} \lesssim {7}\times 10^{-2})$ over the complete energy range.

\begin{figure}[h]
\includegraphics[width=\columnwidth]{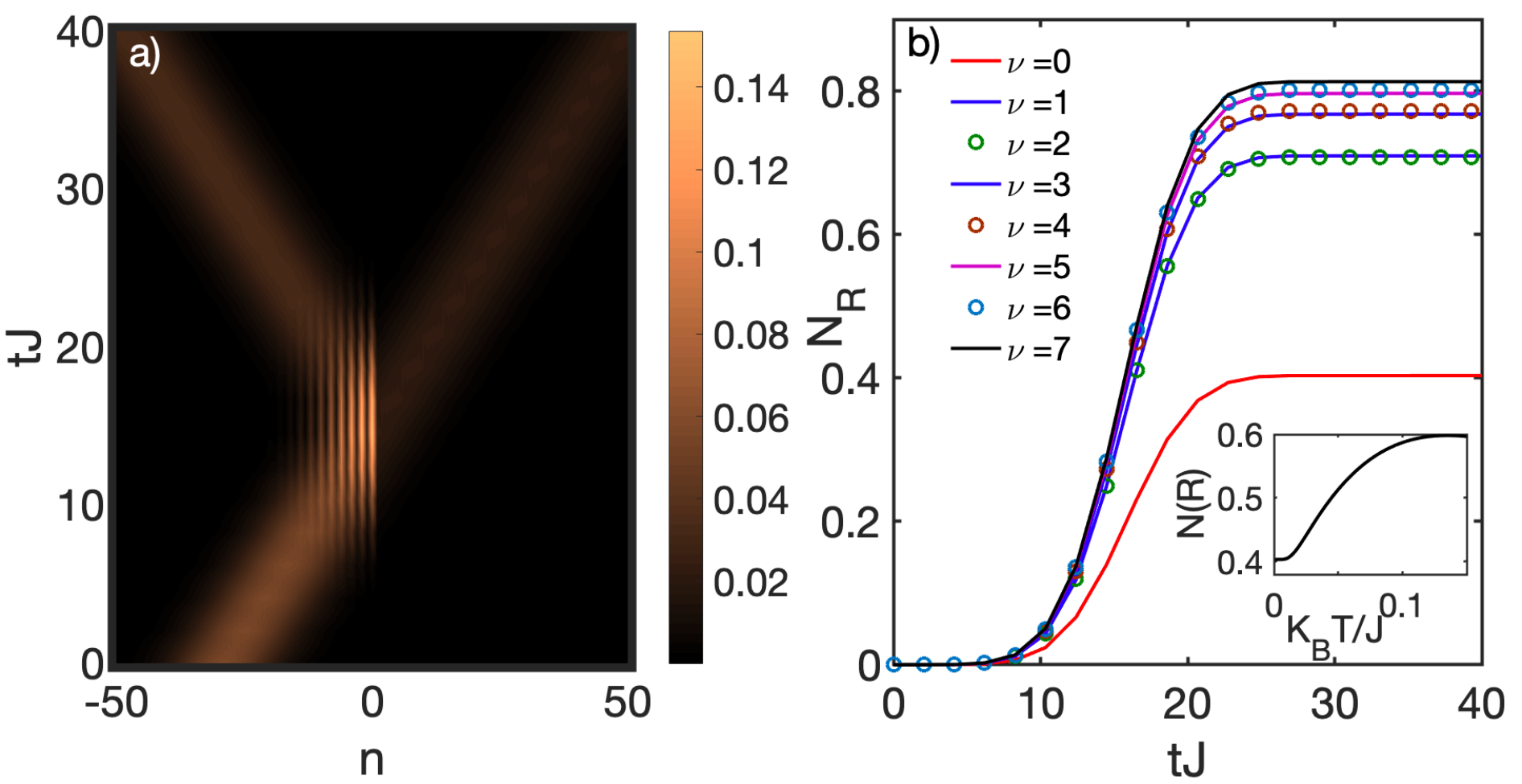}
\caption{Quantum transport enabled by non-adiabatic transitions on $N=100$ sites. (a) Evolution of the excitation probability $p_n(t)=\int d \theta_\alpha |\phi_n(\theta_\alpha,t)|^2$ on the main chain, including vibrations of site $\alpha$. (b) Total probability on the right side of the chain $N_R=\sum_{n>0} p_n(t)$, for the initial harmonic oscillator states $\varphi_\nu(\theta_\alpha)$ of site $\alpha$ and the CU configuration $(\bullet)$ shown in \frefp{spectrum}{a}. If vibrations are frozen as discussed in the text, $N_R\approx 0$. The inset shows transmission as a function of temperature. The dynamics of the full wavefunction including vibrational degrees of freedom is shown in the SI \cite{sup:info} (section II).
}
\label{N_ci}
\end{figure}
Thus, despite the incoming exciton wavepacket's finite energy width of $\sigma_{\cal E}=0.4J$, shown in \frefp{spectrum}{a}, the CU blocks its passage for immobile sites. We verify this using \eref{tdse}, rendering the vibration immobile through a very large frequency $\omega$, while adjusting $M$ to fix the spatial width at $\sigma_\theta\approx 0.2/\pi$ rad, later used for the mobile scenario. 
For such high frequencies, the initial energy is insufficient to excite vibrations and site $\alpha$ remains frozen in the ground state during the transit of the excitation. These immobile simulations exhibit vanishing probability on the right side ($N_R\approx 0$)
for both geometries shown in \frefp{spectrum}{a}.

\begin{figure}[htb]
\includegraphics[width=1\columnwidth]{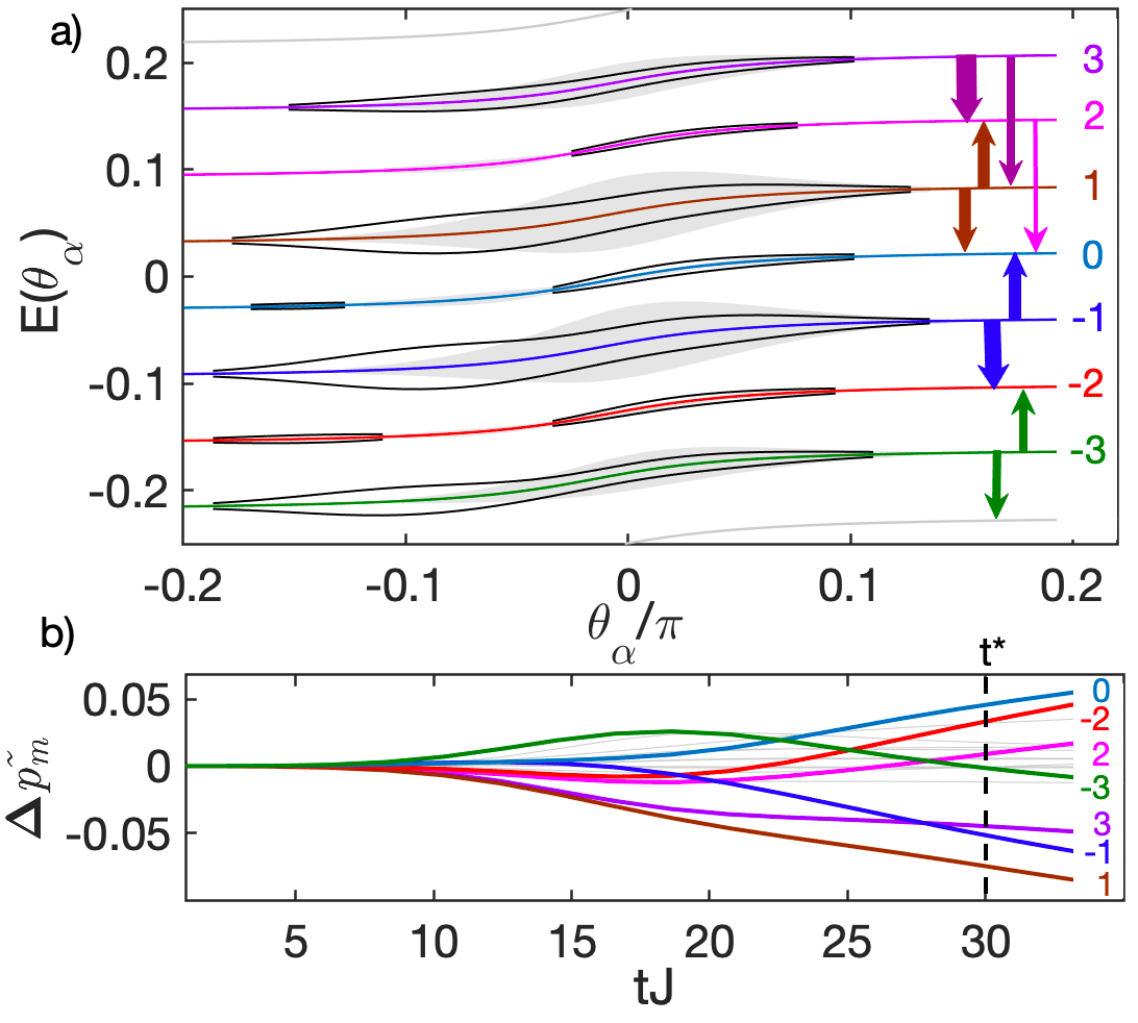}
\caption{Non-adiabatic transitions {cause} transmission. (a) The BO surfaces of $\hat{H}_{el}(\theta_\alpha)$ in the centre of the band for the configuration $(\bullet)$ in \fref{spectrum} are shown as gray or colored lines, with surface index $m$ on the right. Eigenstates with even (odd) indices are dominantly localized on the right (left) of the CU, see text.
 The widths of gray shades from the center line is proportional to the relative density on the surfaces $|\tilde{\phi}_m(\theta)|^2$ at the initial time $t=0$. Black-dashed contours show the same at a later time $t^*$ (after non-adiabatic transitions have occurred), corresponding in (b) to the dashed vertical line. 
All densities are only shown above some cutoff threshold. 
The thickness of vertical arrows between surfaces on the right indicates net non-adiabatic transition probabilities 
until $t^*$ based on $P_{kl}=2\int_0^{t^*} dt \int d\theta_\alpha  \: \mathfrak{Im}[\tilde{\phi}_k^* D_{kl} \tilde{\phi}_l]$  and their colors the surface of origin.
 (b) Net change of population $\Delta \tilde{p}_m = \tilde{p}_m(t)-\tilde{p}_m(0)$ with $\tilde{p}_m(t)=\int d\theta |\tilde{\phi}_m|^2$ on the energy surfaces $m$, colors matching (a). See also supplementary movie \cite{sup:info}.}
\label{pm}
\end{figure}

Transmission becomes significant if site $\alpha$ is unfrozen and increases for higher initial vibrational states $\nu$, as shown in \fref{N_ci}.  
This can not be explained by averaging static transmission coefficients over occupied coordinates $\theta_\alpha$ or assuming adiabatic motion. 
Rather, it can  be directly linked to non-adiabatic transitions at the potential steps in \frefp{spectrum}{b} inherited from the energy surfaces of the CU. To see this, we take a closer look at the relevant band structure in \frefp{pm}{a}, similar to \frefp{spectrum}{b}, with surface index $k=0$ indicating the band center. The interaction of the CU with the main chain effectively cuts the chain in half.  Since the number of sites $N$ is even, this leaves a part with $N/2-1$ sites (left) and another with $N/2$ sites (right), the eigenvalues of which alternate in the middle of the band. The corresponding eigenvectors are thus alternatingly strongly localised on either the left part (odd indices) or the right part (even indices) of the main chain relative to the control-unit. This localization does not qualitatively change for the range of angles $\theta_\alpha$ shown and exceeds $80$\%. For an exciton wavepacket arriving on the left chain, left-localized surfaces are dominantly populated. Without NAC $D_{kl}(\theta_\alpha)$ in \bref{tdse_ad}, this allocation would be preserved, thus prohibiting quantum transport past the CU. However there are non-adiabatic transitions to adjacent energy surfaces, shown by vertical arrows, which modify densities on the BO surfaces at a later time (dotted contours). 
The resulting net populations per surface, shown in (b), indicate that the excitation moves from left localised to right localised surfaces, thus passing the CU.

Transitions to adjacent energy surfaces dominate, as the kinetic energy is less than the surface spacing $|U_{k}-U_{k+1}|$, and NACs $D_{kl}$ decrease for increasing $|k-l|$. Since significant transmission must occur due to increased population on the right localized surfaces that are initially empty, it is solely enabled by non-adiabatic transitions. We confirm this by solving \bref{tdse_ad} with disabled NAC, which shows minimal transmission, about $0.04$\% as in the static scenario. Surface populations continue to change in \frefp{pm}{b} even after the excitation left the CU region, because site $\alpha$ remains in motion.

Since the $D_{kl}$ depend on the mean speed of monomer $\alpha$, which increases with temperature, quantum transport enabled by non-adiabatic transitions could form the basis of temperature sensors in molecular electronics. Within our model we demonstrate this in the inset of \frefp{N_ci}{b}, which shows the temperature dependent transmission  $N_R(T)=\sum_\nu p_\nu N_R(\nu)$, 
where $p_\nu$ is the thermal occupation of vibrational mode $\varphi_{\nu}$ and $N_R(\nu)$ the transmission in that state.

For prominent non-adiabatic transport, our design of amalgamated BO surfaces was crucial. The alternative configuration with grey dashed circles in \frefp{N_ci}{a} shows no transmission, whether mobile or not, since its BO surfaces are flat with respect to $\theta_\alpha$. The most robust way to induce non-adiabaticity into the hybrid structure is a CU that contains conical intersections \cite{yarkony2001conical,worth:CI:review,wuester:CI}, however due to the more technical nature of BO surfaces we discuss this case only in the SI \cite{sup:info} (section III).

\ssection{Conclusion and outlook} We have described a general approach through which potential energy steps as a function of conformation can be controllably induced into the band-structure of a periodic potential, by hybridising it with a flexible control-unit. The latter can block transmission in static or adiabatic cases, while allowing transmission due to non-adiabatic transitions enabled by these steps. Our results can be experimentally tested with cold atoms in optical tweezer arrays \cite{nogrette:hologarrays,wang_tweezerarray} and ultimately might have technological applications for temperature sensors in molecular electronics or molecular transistors and switches involving motional degrees of freedom. The setup constitutes an example where non-adiabaticity is beneficial for excitation transport, complementing adiabatic excitation transport \cite{wuester:cradle,moebius:cradle,leonhardt:switch,ritesh_molagg} where transport is enabled by adiabatic following of quantum eigenstates.  

\acknowledgments
We gladly acknowledge fruitful discussions with Milan \v{S}indelka and Michael Genkin, and SW thanks the Max-Planck society for support under the MPG-IISER partner group program. Also the support and the resources provided by Centre for Development of Advanced Computing (C-DAC) and the National Supercomputing Mission (NSM), Government of India are gratefully acknowledged. {AE}~acknowledges support from the DFG via a Heisenberg fellowship (Grant No EI 872/5-1).

\bibliography{ref}

\begin{thebibliography}{6}
\bibitem[{par()}]{par_SI}
\bibinfo{note}{{\it For the configuration with CI:} The radius of the circle is
  $R = 0.21\delta$, $a$ = $1.3 \times \sqrt{3} R$, with distances defined in
  Fig.~1 of the main article. $\theta_{\beta\alpha}=\theta_{\eta\beta}=2\pi/3$
  and monomer onsite energies in the ring are $E_n=-20 J$.}
\end{thebibliography}
\end{document}


 \title{Supplementary Material: Quantum transport enabled by non-adiabatic transitions}
\author{Ajith Ramachandran}
\affiliation{Department of Physics, Indian Institute of Science Education and Research, Bhopal, Madhya Pradesh 462 066, India}
\affiliation{Department of Physics, Christ College, Irinjalakuda, Kerala 680125, India}
\author{Alexander Eisfeld}
\affiliation{Max Planck Institute for the Physics of Complex Systems, N\"othnitzer Stra{\ss}e 38, D-01187 Dresden, Germany}
\author{Sebastian W\"uster}
\email{sebastian@iiserb.ac.in}
\affiliation{Department of Physics, Indian Institute of Science Education and Research, Bhopal, Madhya Pradesh 462 066, India}
\author{Jan-Michael Rost}
\affiliation{Max Planck Institute for the Physics of Complex Systems, N\"othnitzer Stra{\ss}e 38, D-01187 Dresden, Germany}
\vskip0.2cm

\begin{abstract}
This supplemental material provides details regarding the dynamics of the wavefunction, non-adiabatic couplings and an alternative configuration with a conical intersection in the BO surfaces.
\end{abstract}
\beginsupplement

\maketitle


 \section{Non-adiabatic coupling} \label{nac}
When writing the Schr{\"o}dinger equation in the adiabatic basis, Eq.~(4) of the main text, one encounters the non-adiabatic coupling terms
\begin{align} \label{nonadc}
D_{kl}(\theta_\alpha)= &-\frac{1}{2M}(\braket{\psi_k(\theta_\alpha)|\boldsymbol{\nabla}^2|\psi_l(\theta_\alpha)}\CR
&+2 \braket{\psi_k(\theta_\alpha)|\boldsymbol{\nabla}|\psi_l(\theta_\alpha)}\cdot \boldsymbol{\nabla}).
\end{align}
Note that this is still a differential operator that acts on the BO surface wavefunction $\tilde{\phi}_l(\theta_\alpha,t)$ in Eq.~(4).
The second term typically dominates, and is proportional to $\boldsymbol{\nabla}\tilde{\phi}_l(\theta_\alpha,t)$, which is connected to the velocity of nuclear motion. Faster motion implies stronger non-adiabatic effects. We can rewrite the two quantum matrix elements using the Hellman-Feynman theorem \cite{hellmann1939einfuhrung,politzer2018hellmann}, 
to see that they scale as $|U_k-U_l |^{-1}$. Thus non-adiabatic effects will become large near avoided crossings or conical intersections.

We use \eref{nonadc} for Fig. 4 in the main article to calculate the probability of non-adiabatic transitions.
From the TDSE in Eq. (4) of the main text, one can derive the evolution equation
\begin{align} \label{surfpopchange}
\dot{\tilde{p}}_k = 2\int d\theta_\alpha\:  \sum_{\ell \neq k}\mathfrak{Im}[\tilde{\phi}_k(\theta_\alpha)^* D_{kl}(\theta_\alpha) \tilde{\phi}_\ell(\theta_\alpha)] 
\end{align}
for the net population $\tilde{p}_k=\int d\theta_\alpha\: |\tilde{\phi}_k(\theta_\alpha)|^2$ on the adiabatic surface $k$.

Based on that we interpret the quantity $P_{k\ell}$ given in the caption of Fig. 4 as transition rate from surface $\ell$ to surface $k$ if positive, and the reverse if negative.

 \section{Configuration with Conical Intersection in BO Surfaces} \label{CI}
In the main article we amalgamate the band structure of the chain with the slanting BO surfaces present in a trimer side-unit for simplicity of presentation.
However parameters have to be carefully adjusted to generate the desired surfaces in this case. We found that a geometry where avoided crossings or potential steps are robustly inherited in 
the hybrid structure, is when the isolated CU possessed a conical intersection (CI) in its spectrum \cite{yarkony2001conical,worth:CI:review}. For the present model, this occurs when the trimer sites form an equilateral triangle \cite{wuester:CI}, which we consider here with parameters given in footnote \cite{par_SI}. The transmission profile corresponding to this configuration possesses a single Fano resonance dip at ${\cal E}=0 J$. 

\begin{figure}[H]
\includegraphics[width=\columnwidth]{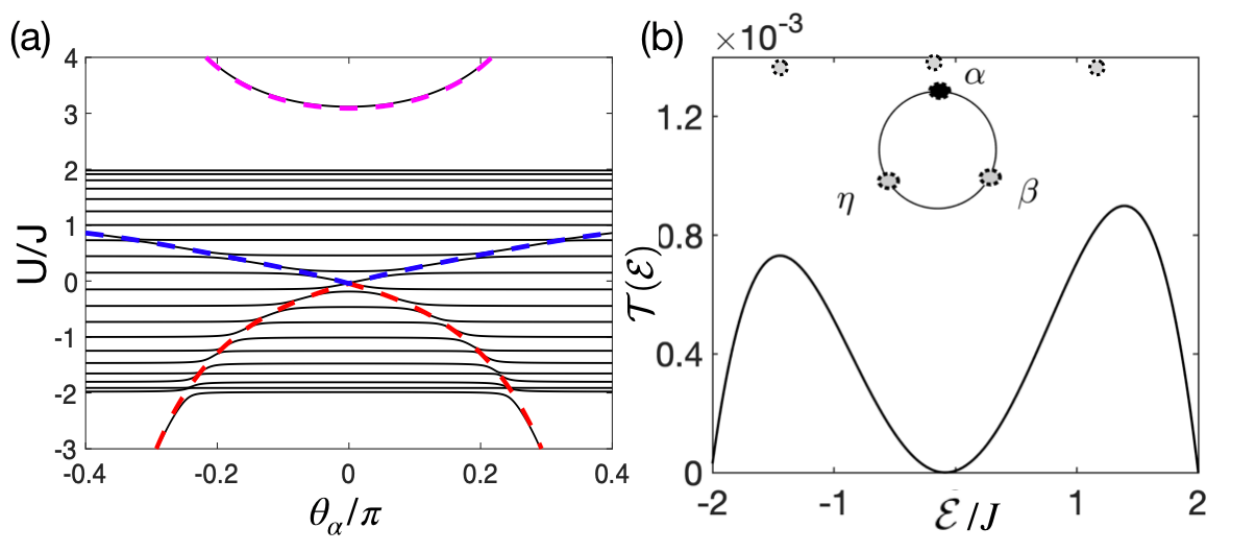}
\caption{(a) Spectrum (black lines) of the electronic Hamiltonian $\sub{\hat{H}}{el}$ in Eq. (2) for a chain of $N=20$ sites that is hybridized with a trimer control-unit creating a CI. We consider monomer $\alpha$ mobile on a ring, with polar angle $\theta_\alpha$. Colored lines are the energies $U_j^{(C)}$ of the trimer control-unit without chain. For other parameters and geometry, see \cite{par_SI}. (b) Very small transmission ${\cal T}$ past the control-unit (note the scaling by $10^{-3}$) as function of incoming energy $\cal E$ of the excitation, for the configuration shown in the inset. All monomers are immobile here. 
}
\label{ci}
\end{figure}

\begin{figure}[H]
\includegraphics[width=\columnwidth]{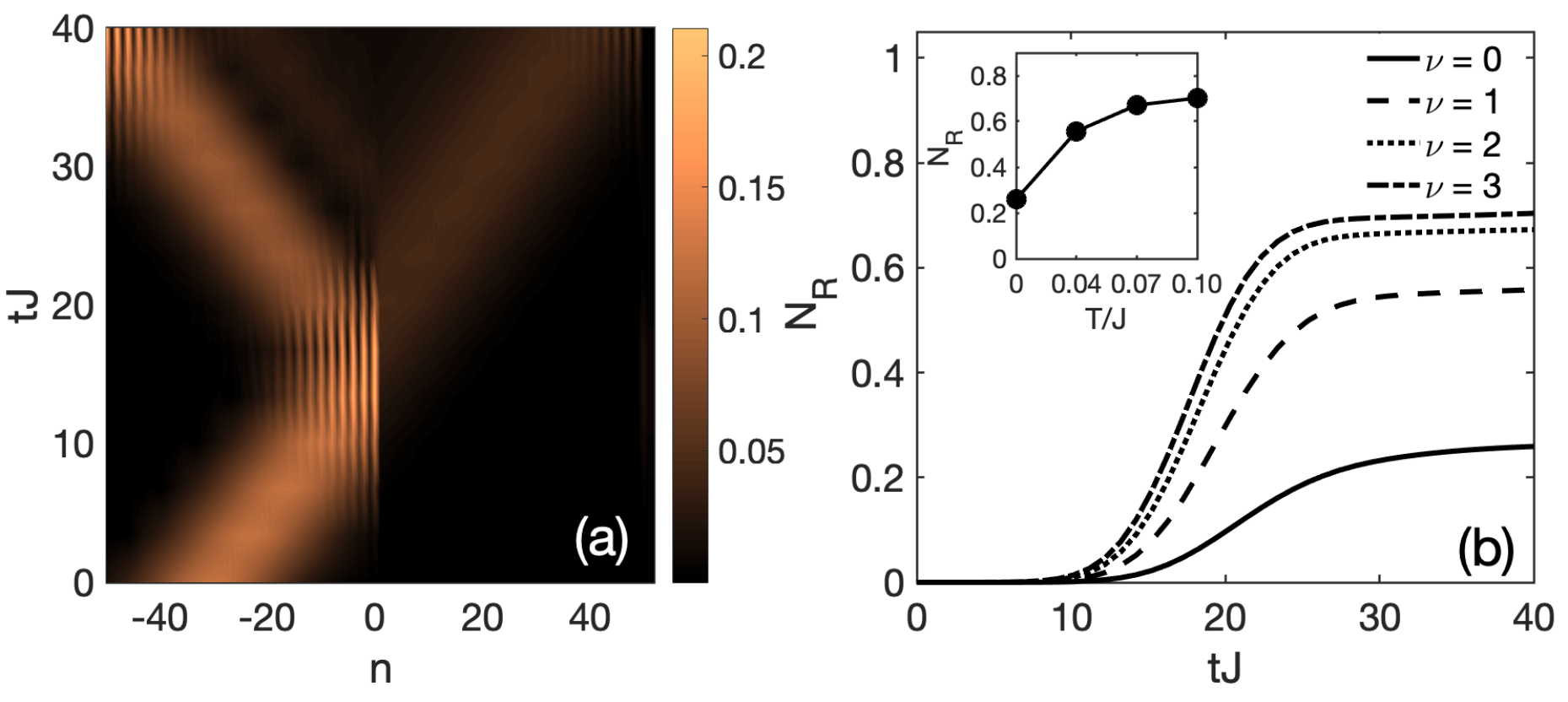}
\caption{Quantum transport enabled by non-adiabatic transitions, exploiting a CI in the CU. (a) Evolution of the excitation probability $p_n(t)=\int d \theta_\alpha |\phi_n(\theta_\alpha,t)|^2$ on the main chain, including vibrations of site $\alpha$. (b) Total probability on the right side of the chain $N_R=\sum_{n>0} p_n(t)$, for the initial harmonic oscillator states $\varphi_\nu(\theta_\alpha)$ of monomer $\alpha$ and the solid dot configuration in \frefp{ci}{b}. If vibrations are frozen, $N_R\approx 0$. The inset shows transmission as a function of temperature.}
\label{N_ci}
\end{figure}

The BO surfaces for this configuration exhibit a CI as shown in \fref{ci}(a). 
\fref{N_ci}(a) and (b) show that the motion of monomer $\alpha$ introduces a nonzero transmission at  ${\cal E}=0J$. Although only the left surfaces are populated initially, the nonadiabatic dynamics populates the right surfaces as time progresses leading to the possibility of nonzero transmission. Hence, the non-adiabatic features due to CI can also enhance the transmission beyond the scatterer.

%

 \section{Dynamics of the full wavefunction} \label{dyns}
\begin{figure}[h]
\includegraphics[width=\linewidth]{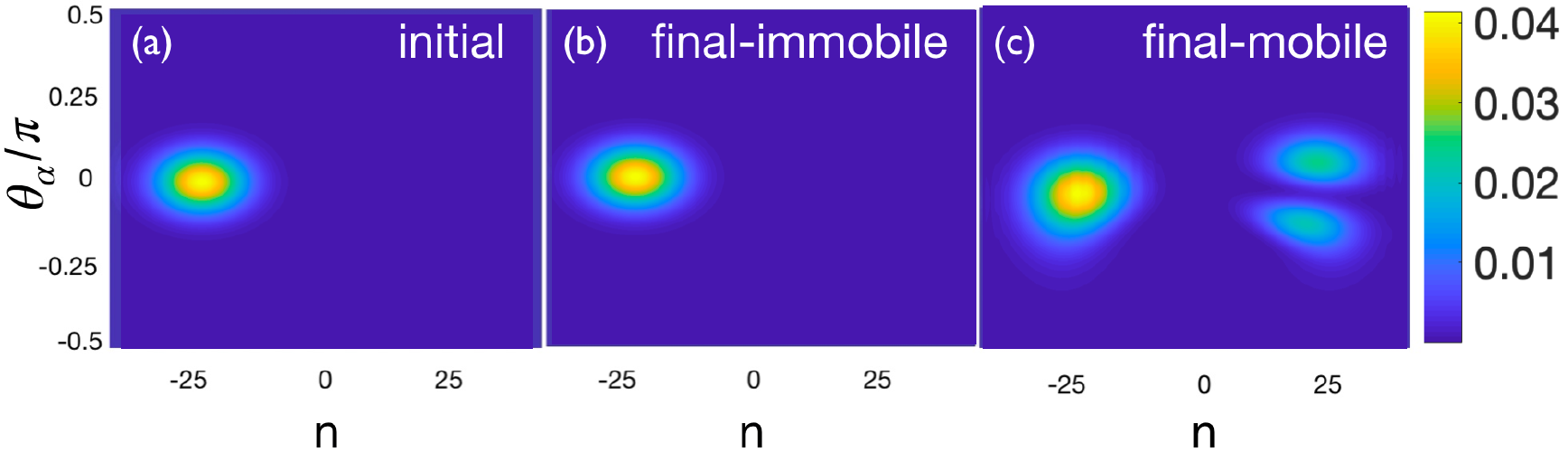}
\caption{\label{sketch0} Combined electronic and vibrational dynamics of the hybrid system. (a) 
Probability density $|\phi_n(\theta_\alpha,t)|^2$ of (a) the
incident wavepacket, (b) wavepacket after scattering for the case when the monomers are static, (c) wavepacket after scattering for the case when monomer $\alpha$ may move.
}
\label{dyn} 
\end{figure}

 In the main text, we have primarily focussed on the quantum transport of the excitation. 
 The propagation of the wavefunctions $\phi_n(\theta_\alpha,t)$ of course also gives access to the vibrational dynamics of monomer $\alpha$, which we discuss here. 
 The complete probability density $|\phi_n(\theta_\alpha,t)|^2$ in terms of excitation location $n$ and angle of the vibrating monomer alpha $\theta_\alpha$ propagated by Eq.~(3) of the main text is shown in \fref{dyn} for the configuration with CI. \fref{dyn}(a) shows the incident wavepacket at $t=0$. \fref{dyn}(b) shows the wavepacket after the scattering when the monomer is static. It can be seen that the wavepacket remains on the left part of the chain indicating a complete suppression of the transmission beyond the scatterer. \fref{dyn}(c) shows the wavepacket after the scattering when the monomer is mobile. Due to the slope of BO surfaces visible in Fig.~4 in the main article, it then also does begin to move, which increases the spatial spread in the wavepacket. 
%
 \section{Caption of supplementary movie} \label{movie_caption}
The top panel shows the instantaneous excitation probability $p_n(t)$ of main chain monomer $n$ at the time $t$ shown. The CU is attached at $n=0$ indicated by the blue vertical line and probabilities thereon are not shown. The bottom panel shows the surface densities $|\tilde{\phi}_m(\theta)|^2$ in a similar style
 as Fig.~4, as gray shade surrounding BO surfaces shown as colored lines.
 \bibliography{ref}